\documentstyle[12pt]{article}
\begin{document}
\thispagestyle{empty}
\begin{center}{DERIVATION OF THE TULLY-FISHER LAW FROM GENERAL RELATIVITY THEORY: DOUBTS
ABOUT THE EXISTENCE OF HALO DARK MATTER}\end{center}
\vspace{1cm}
\begin{center}{Silvia Behar and Moshe Carmeli}\end{center} 
\begin{center}{Department of Physics, Ben Gurion University, Beer Sheva 84105, 
Israel}\end{center}
\begin{center}{(E-Mail: carmelim@bgumail.bgu.ac.il)}\end{center}
\vspace{2cm} 
\begin{center}{ABSTRACT}\end{center}
Observations show that for disk galaxies the fourth power of the circular 
velocity $v_c^4$ of stars around the core of the galaxy is proportional to
the luminosity $L$, $v_c^4\propto L$. This is known as the Tully-Fisher law.
Since $L$ is proportional to the mass $M$ of the galaxy, it follows that 
$v_c^4\propto M$. Newtonian mechanics, however, yields $v_c^2=GM/r$ for a
circular motion. In order to rectify this big difference, astronomers assume 
the existence of dark matter. In this paper we show that general relativity 
theory yields a term of the form $v_c^4\propto GMc/\tau$, where $G$ is 
Newton's gravitational constant, $c$ is the speed of light and $\tau$ the 
Hubble time. This puts doubts about the necessity and existence of halo dark 
matter for galaxies.
\newpage
\section{Introduction}
The universe is observed through electromagnetic waves. From the stars the
waves are visible, from hot plasmas they are x-rays, from the hyperfine 
transition in hydrogen they are radio waves, and they are microwaves from the 
cosmic background radiation.

Not all matter in the universe, however, emits detectable radiation. Examples
of this are blackholes and zero-mass neutrinos. The difference between the
detectable mass and the total mass that should be according to the laws of 
gravity is ascribed to the so-called dark matter, whose existence is inferred
only from its gravitational interaction. 

The visible parts of the galaxies are composed mainly of stars which do not
satisfy Newton's mechanics and thus are hypothized to be surrounded by 
extended halos of dark matter which may be a factor of 30 or more in both
mass and size. The existence of the planet Neptune was predicted from the
unexpected residuals in the motion of Uranus.

A negative example, on the other hand, is the precession of the planet 
Murcury's perihelion. A hypothetical planet or a ring of matter, inside Murcury's
orbit, was hypothized to exist in order to explain the anomaly. No planet or
material ring was observed.

As is well known, the anomaly was resolved by Einstein's general relativity 
theory [1]. This is a reminder that much of the assumed missing matter might
be explained by new theories.

In this paper we use a space-velocity version of general relativity theory [2-4]
in combination with Einstein's standard general relativity theory to show that 
much of the unexplained observation can be satisfactorily described. More 
precisely, we simply prove that the Tully-Fisher law is included in the 
equations of motion obtained.
\section{The Tully-Fisher Law} 
Astronomical observations show that for disk galaxies the fourth power of the
circular speed of stars moving around the core of the galaxy, $v_c^4$, is
proportional to the total luminosity $L$ of the galaxy to an accuracy of more 
than two orders of magnitude in $L$, namely $v_c^4\propto L$. Since $L$ is
proportional to the mass $M$ of the galaxy, one obtains $v_c^4\propto M$. 
This is known as the Tully-Fisher law [5]. No dependence on the distance of
the star from the center of the galaxy as Newton's law $v_c^2=GM/r$ requires
for circular motion.

In order to rectify this deviation from Newton's laws astronomers assume the
existence of halos around the galaxy which are filled with dark matter and
arranged in such ways so as to satisfy the Tully-Fisher law for each 
particular situation.

It is well known that Newton's second law also follows from Einstein's general 
relativity theory in the lowest approximation in $v/c$, where $v$ is a 
characteristic velocity and $c$ the speed of light. For this reason we exclude
the possibility of modifying Newton's second law of motion by such as adding 
to it a term which takes care of the anomaly [6,7]. Any arbitrary modification
of Newton's second law is therefore spurious even if it yields results that
fit observations quite well.
\section{Constraints on Motion of Stars}
A star moving around the galaxy experiences the expansion of the universe.
This is a constraint on the dynamical system that should be taken into account,
and without which the theory is invalid.

The expansion of the universe causes an increase to the distance between the 
star and the center of the galaxy. But when this distance increases, the 
circular velocity changes accordingly. This constraint on the dynamical
system should be taken into account along with the centrifugal formula 
$v_c^2=GM/r$. But a space-velocity version of general relativity is exactly
the right theory to give this extra relation between the mass of the galaxy, 
the circular velocity and the distance of the star to the center of the galaxy.

In this paper we derive the extra relation from the space-velocity version of
general relativity theory and show that its combination with Einstein's 
general relativity yield a term of the form $v_c^4\propto M$.
\section {Equations of Motion} 
In Einstein's general relativity theory the equations of motion follow from
the vanishing of the covariant divergence of the energy-momentum tensor. This
is a result of the restricted Bianchi identities. The equations obtained are
usually the geodesic equation. By means of a successive approximation in
$v/c$, one obtains the Newtonian equations of motion and its generalization
to a higher accuracy [8-25].

In the space-velocity version of general relativity the situation is the same.
Accordingly one has
$$\frac{d^2x^\mu}{ds^2}+\Gamma_{\alpha\beta}^\mu\frac{dx^\alpha}{ds}
\frac{dx^\beta}{ds}=0.\eqno(1)$$ 
We now find the lowest approximation of Eq. (1) in terms of $t/\tau$, where 
$t$ is a characteristic cosmic time and $\tau$ is the Hubble time, using the
Einstein-Infeld-Hoffmann method [9,10].

To this end we change variables in Eq. (1) from $s$ to $v$, where $v$ is 
related to the velocity-like coordinate $x^0$ by $x^0=\tau v$. A simple
calculation gives
$$\frac{d^2x^k}{dv^2}+\left(\Gamma_{\alpha\beta}^k-\frac{1}{\tau}
\frac{dx^k}{dv}\Gamma_{\alpha\beta}^0\right)\frac{dx^\alpha}{dv}
\frac{dx^\beta}{dv}=0,\eqno(1a)$$ 
with $k=1,2,3$. As seen, one can neglect the second term in the paranthesis
since it is one order smaller than the first, and thus
$$\frac{d^2x^k}{dv^2}+\Gamma_{\alpha\beta}^k\frac{dx^\alpha}{dv}
\frac{dx^\beta}{dv}\approx 0.$$
The second term is equal to 
$$\Gamma_{00}^k\left(\frac{dx^0}{dv}\right)^2+2\Gamma_{0b}^k\frac{dx^0}{dv}
\frac{dx^b}{dv}+\Gamma_{ab}^k\frac{dx^a}{dv}\frac{dx^b}{dv}.$$
But $x^0=\tau v$, thus the 2nd and the 3rd terms may be neglected with 
respect to the 1st, and we obtain
$$\frac{d^2x^k}{dv^2}+\tau^2\Gamma_{00}^k\approx 0.$$

The Christoffel symbol can be calculated also,
$$\Gamma_{00}^k=\frac{1}{2}g^{k\rho}\left(2\partial_0g_{\rho 0}-\partial_\rho
g_{00}\right).$$
Again a $x^0$-derivative $\partial_0=\tau^{-1}\partial_v$ which is of higher 
order in $t/\tau$, thus 
$$\Gamma_{00}^k\approx -\frac{1}{2}g^{k\rho}\partial_\rho g_{00}\approx
-\frac{1}{2}g^{ks}\partial_s g_{00}.$$
Since $g^{ks}\approx \eta^{ks}=-\delta^{ks}$, we obtain
$$\Gamma_{00}^k\approx\frac{1}{2}\frac{\partial g_{00}}{\partial x^k},\eqno(2)$$
and thus the geodesic equation yields
$$\frac{d^2x^k}{dv^2}+\frac{\tau^2}{2}\frac{\partial g_{00}}{\partial x^k}
\approx 0.\eqno(3)$$

Writing now $g_{00}=1+2\phi/\tau^2$, we then obtain
$$\frac{d^2x^k}{dv^2}=-\frac{\partial\phi}{\partial x^k},\eqno(4)$$
for the equations of motion in the lowest approximation. It remains to find
out the function $\phi\left(x\right)$.
\section{Field Equations}  
To find out the function $\phi$ we have to solve the gravitational field 
equations. Again this should be in the lowest approximation as explained 
before. We have [2-4]
$$R_{\mu\nu}=\kappa\left(T_{\mu\nu}-\frac{1}{2}g_{\mu\nu}T\right),\eqno(5)$$
where $T=T_{\rho\sigma}g^{\rho\sigma}$ and 
$$\kappa=\frac{8\pi k}{\tau^4},\eqno(6)$$
with $k=G\left(\tau^2/c^2\right)$. We then have
$$T=T_{\mu\nu}g^{\mu\nu}\approx T_{\mu\nu}\eta^{\mu\nu}\approx T_{00}\eta^{00}
=T_{00}.\eqno(7)$$
Thus we obtain
$$R_{00}=\kappa\left(T_{00}-\frac{1}{2}g_{00}T\right)\approx\frac{1}{2}\kappa
T_{00}=\frac{1}{2}\kappa\tau^2\rho\left(x\right),\eqno(8)$$
where $\rho\left(x\right)$ is the mass density.

The approximate value of $R_{00}$ is
$$R_{00}=\frac{\partial\Gamma_{00}^\rho}{\partial x^\rho}-
\frac{\partial\Gamma_{0\rho}^\rho}{\partial x^0}+\Gamma_{00}^\rho
\Gamma_{\rho\sigma}^\sigma-\Gamma_{0\rho}^\sigma\Gamma_{0\sigma}^\rho\approx
\frac{\partial\Gamma_{00}^\rho}{\partial x^\rho}\approx
\frac{\partial\Gamma_{00}^s}{\partial x^s}.\eqno(9)$$
Using now Eq. (2) for the value of the Christoffel symbol, we obtain
$$R_{00}\approx\frac{\partial\Gamma_{00}^s}{\partial x^s}=\frac{1}{2}
\frac{\partial^2g_{00}}{\partial x^s\partial x^s}=\frac{1}{2}\nabla^2g_{00}
\approx\frac{1}{\tau^2}\nabla^2\phi,\eqno(10)$$
where $\nabla^2$ is the ordinary Laplace operator.

Equations (8) and (10) then give
$$\nabla^2\phi=\frac{1}{2}\kappa\tau^4\rho,\eqno(11)$$
or, using Eq. (6),
$$\nabla^2\phi\left(x\right)=4\pi k\rho\left(x\right).\eqno(12)$$
This equation is exactly the Newtonian equation for gravity but with 
$k=G\tau^2/c^2$ replacing the Newtonian constant $G$ [2-4].
\section{Integration of the Equations of Motion}
The integration of the equation of motion (4) is identical to that familiar
in classical Newtonian mechanics. But there is an essential difference which
should be emphasized.

In Newtonian equations of motion one deals with a path of motion in the 
3-space. In our theory we do not have that situation. Rather, the paths here
indicate locations of particles in the sense of the Hubble distribution which
now takes a different physical meaning. With that in mind we proceed as 
follows.

Equation (4) yields the first integral
$$\left(\frac{ds}{dv}\right)^2=\frac{kM}{r},\eqno(13)$$
where $v$ is the velocity of the particles, in analogy to the  Newtonian
$$\left(\frac{ds}{dt}\right)^2=\frac{GM}{r}.\eqno(14)$$ 
In these equations $s$ is the length parameter along the path of accumulation
of the particles.

Comparing Eqs. (13) and (14), and remembering that $k=G\tau^2/c^2$, we obtain
$$\frac{ds}{dv}=\frac{\tau}{c}\frac{ds}{dt},\eqno(15)$$
thus
$$\frac{dv}{dt}=\frac{c}{\tau}.\eqno(16)$$
Accordingly we see that the particle experiences an acceleration $a_0=c/\tau=
cH_0$ directed outward when the motion is circular. This extra term is not
included or derivable from Einstein's general relativity theory and it appears
only in ours.
\section{Effective Potential}
The motion of a particle in a central field is best described in terms of an
``effective potential", $V_{eff}$. In Newtonian mechanics this is given by 
[26]
$$V_{eff}=-\frac{GM}{r}+\frac{L^2}{2r^2},\eqno(17)$$
where $L$ is the angular momentum per mass unit. In our case the effective
potential is
$$V_{eff}\left(r\right)=-\frac{GM}{r}+\frac{L^2}{2r^2}+a_0r.\eqno(18)$$

The circular motion is obtained at the minimal value of (18), i.e.
$$\frac{dV_{eff}}{dr}=\frac{GM}{r^2}-\frac{L^2}{r^3}+a_0=0,\eqno(19)$$
with $L=v_cr$, and $v_c$ is the rotational velocity. This gives
$$v_c^2=\frac{GM}{r}+a_0r,\eqno(20)$$
thus
$$v_c^4=\left(\frac{GM}{r}\right)^2+2GMa_0+a_0^2r^2,\eqno(21)$$
where $a_0=c/\tau=cH_0$.
\section{Concluding Remarks}
The first term on the right-hand side of Eq. (21) is purely Newtonian which
cannot be evaded by any reasonable theory. The second one is the Tully-Fisher
term. The third term is extremely small at the range of distances of stars 
around a galaxy.

It has been shown [6,7] that a term of the form $GMa_0=GMcH_0$ can explain
most of the observations of the dynamics of stars around the galaxies. The
``modified Newtonian law of motion" proposed in Refs. 6 and 7 was by adding
arbitrarily an attractive force term in the very far distances. We have seen, 
however, that our theory predicts a repulsive force term rather than an
attractive one. 

In conclusion it appears to us that one should have great doubts about the 
necessity and existence of halo dark matter around galaxies. Rather, this is
a property of spacetime [27,28].
\newpage
\section*{References}
1. See, for example, M. Carmeli, {\it Classical Fields: General Relativity
and Gauge Theory} (Wiley, New York, 1982).\newline
2. M. Carmeli, {\it Commun. Theor. Phys.} {\bf 5}, 159 (1996).\newline 
3. M. Carmeli, {\it Commun. Theor. Phys.} {\bf 6}, 45 (1997).\newline
4. M. Carmeli, {\it Commun. Math. Theor. Phys.} {\bf 1}, 50 (1998).\newline
5. B.C. Whitemore, Rotation curves of spiral galaxies in clusters, in: {\it
Galactic Models}, Eds. J.R. Buchler, S.T. Gottesman and J.H. Hunter, Jr.
(Ann. New York Academy Sciences, Vol. 596, New York, 1990).\newline
6. M. Milgrom, {\it Astrophys. J} {\bf 270}, 365, 371, 384 (1983).\newline
7. R.H. Sanders, {\it Astron. Astrophys. Rev.} {\bf 2}, 1 (1990).\newline
8. A. Einstein and J. Grommer, {\it Preuss. Akad. Wiss., Phys.-Math. Klasse}
{\bf 1}, 2 (1927).\newline
9. A. Einstein, L. Infeld and B. Hoffmann, {\it Annals of Mathematics} 
{\bf 39}, 65 (1938).\newline
10. A. Einstein and L. Infeld, {\it Can. J. Math.} {\bf 1}, 209 (1949).\newline
11. L. Infeld and A. Schild, {\it Revs. Mod. Phys.} {\bf 21}, 408 (1949).\newline
12. L. Infeld, {\it Revs. Mod. Phys.} {\bf 29}, 398 (1957).\newline 
13. V. Fock, {\it Revs. Mod. Phys.} {\bf 29}, 325 (1957).\newline
14. V. Fock, {\it The Theory if Space, Time and Gravitation} (Pergamen Press, 
Oxford, 1959).\newline
15. L. Infeld and J. Plebanski, {\it Motion and Relativity} (Pergamen Press,
Oxford, 1960).\newline
16. B. Bertotti and J. Plebanski, {\it Ann. Phys. (N.Y.)} {\bf 11}, 169 
(1960).\newline
17. M. Carmeli, {\it Phys. Lett.} {\bf 9}, 132 (1964).\newline
18. M. Carmeli, {\it Phys. Lett.} {\bf 11}, 24 (1964).\newline 
19. M. Carmeli, {\it Ann. Phys. (N.Y.)} {\bf 30}, 168 (1964).\newline
20. M. Carmeli, {\it Phys. Rev.} {\bf 138}, B1003 (1965).\newline
21. M. Carmeli, {\it Nuovo Cimento} {\bf 37}, 842 (1965).\newline
22. M. Carmeli, {\it Ann. Phys. (N.Y.)} {\bf 34}, 465 (1965).\newline
23. M. Carmeli, {\it Ann. Phys. (N.Y.)} {\bf 35}, 250 (1965).\newline
24. M. Carmeli, {\it Phys. Rev.} {\bf 140}, B1441 (1965).\newline
25. T. Damour, Gravitational radiation and the motion of compact bodies, in
{\it Gravitational Radiation}, N. Deruelle and T. Piran, Eds. (North-Holland,
Amsterdam 1983), pp. 59-144.\newline
26. H. Goldstein, {\it Classical Mechanics} (Addison-Wesley, Reading, Mass.,
1980).\newline
27. M. Carmeli, {\it Inter. J. Theor. Phys.} {\bf 37}, 2621 (1998).\newline
28. M. Carmeli, {\it Commun. Math. Theor. Phys.} {\bf 1}, 54 (1998).
\end{document}